%% file: main_IEEE.tex
\documentclass[conference]{IEEEtran}
\usepackage{amsmath,amssymb,amsfonts}
\input{cops}

\IEEEoverridecommandlockouts
\usepackage{cite}
\usepackage{graphicx}
\usepackage{textcomp}
\usepackage{xcolor}
\usepackage{algorithm}
\usepackage{algpseudocode}
\usepackage{booktabs}
\usepackage{multirow}
\usepackage{threeparttable}

\begin{document}

\title{OriginPruner: Leveraging Method Origins for Guided Call Graph Pruning}


\author{\IEEEauthorblockN{Amir M. Mir}
\IEEEauthorblockA{\textit{Software Technology Department} \\
\textit{Delft University of Technology}\\
Delft, The Netherlands \\
s.a.m.mir@tudelft.nl}
\and
\IEEEauthorblockN{Mehdi Keshani}
\IEEEauthorblockA{\textit{Software Technology Department} \\
\textit{Delft University of Technology}\\
Delft, The Netherlands \\
m.keshani@tudelft.nl}
\and
\IEEEauthorblockN{Sebastian Proksch}
\IEEEauthorblockA{\textit{Software Technology Department} \\
\textit{Delft University of Technology}\\
Delft, The Netherlands \\
s.proksch@tudelft.nl}
}

\maketitle
\begin{abstract}
Most static program analyses depend on Call Graphs (CGs), including reachability of security vulnerabilities.
Static CGs ensure soundness through over-approximation, which results in inflated sizes and imprecision.
Recent research has employed machine learning (ML) models to prune false edges and enhance CG precision.
However, these models require real-world programs with high test coverage to generalize effectively and the inference is expensive.
In this paper, we present \tool{OriginPruner}, a novel call graph pruning technique that leverages the method origin, which is where a method signature is first introduced within a class hierarchy.
By incorporating insights from a localness analysis that investigated the scope of method interactions into our approach, OriginPruner confidently identifies and prunes edges related to these origin methods.
Our key findings reveal that
(1) dominant origin methods, such as \code{Iterator.next}, significantly impact CG sizes; (2) derivatives of these origin methods are primarily local, enabling safe pruning without affecting downstream inter-procedural analyses;
(3) \tool{OriginPruner} achieves a significant reduction in CG size while maintaining the soundness of CGs for security applications like vulnerability propagation analysis; and
(4) \tool{OriginPruner} introduces minimal computational overhead.
These findings underscore the potential of leveraging domain knowledge about the type system for more effective CG pruning, offering a promising direction for future work in static program analysis.
\end{abstract}


\section{Introduction}
Call graphs (CGs) are an essential element of various static program analyses, including performance profiling or security vulnerability assessment.
Call graphs representing function invocations within programs~\cite{ryder1979constructing, callahan1990constructing}.
Creating this representation poses a significant challenge, even for small-scale programs~\cite{ali2012application}, as it is necessary to achieve a balance between \emph{soundness}, i.e., ensuring no legitimate function calls are missed, and \emph{precision}, i.e., avoiding superfluous calls.
Despite recent initiatives towards more practical static analyses~\cite{keshani2024frankenstein}, this trade-off is usually decided in favor of soundness, and CGs are typically over-approximated and imprecise~\cite{utture2022striking, antal2023javascript, mir2024effectiveness}.

This leads to a major issue, especially in object-oriented programming languages, where the scalability of CGs suffers from ubiquitous subtyping.
For every call site in a program, a CG generator has to identify all possible target types through a \emph{class-hierarchy analysis}.
For example, a single call to the target method \code{C.m} will be expanded by adding edges to \emph{all} overridden variants of \code{m} that have been created in subtypes of \code{C}.
Cases like \code{Object.toString} add an enormous \emph{fan out} to every call-graph node that contains such a call, which severely limits the scalability of analyses.
As a result, CGs are very big, which affects the performance of downstream static analyses.
Research on static analysis tries to improve the precision by pruning unreachable types, for example, using enhanced pointer analysis based on context-sensitivity or flow-sensitivity~\cite{bravenboer2009strictly, mangal2015user, tan2016making}.
However, such techniques are very computationally expensive even for small improvements~\cite{mir2023effect}.

Recent efforts have introduced novel machine-learning-based pruning techniques that can reduce the size of call graphs by eliminating false edges, e.g., \tool{CGPruner}~\cite{utture2022striking} and \tool{AutoPruner}~\cite{le2022autopruner}.
These approaches learn to prune false CG edges from dynamic traces of actual program executions.
\tool{CGPruner} uses structural features while \tool{AutoPruner} learns from a combination of structural and semantic features.
ML-based CG pruning approaches greatly enhance CG precision by up to 45\%, though they cause a substantial loss in the recall/soundness of call graphs, which renders these approaches impractical for security-focused applications like vulnerability propagation.
To alleviate this, Mir et al.~\cite{mir2023effect} conducted an empirical study on the effectiveness of ML-based CG pruning approaches and they proposed using a conservative strategy that trades a slight loss in recall for a much higher precision. 

While ML-based CG pruning approaches offer promising results, we believe that their quality is conceptually limited by the choice of features that are used to model the CG.
In this paper, we will approach the pruning problem from a different perspective.
In contrast to existing approaches that are based on basic graph features (like a nodes \emph{in-degree}) or \emph{pseudo-semantic} features (source code as plain text), we propose to base the pruning decision on domain knowledge about the type system.
We will investigate in this paper, whether a novel CG-pruning approach can use \emph{actual} semantic knowledge about the type system to improve its pruning decisions.
The idea is rooted in \emph{object-oriented programming}, where methods can be overridden in subclasses.
We locate the first definition of each unique method signature, the \emph{origin method}, and find that, in practice, a small number of origin methods like \code{Object.hashCode} or \code{Iterator.next} are frequently subtyped and used so often throughout typical code bases that they are responsible for a large fraction of the overall size of the resulting CG.
We believe that instead of investigating general CG pruners, it is more promising to identify and prune only those problematic edges.
We evaluate our intuition through four research questions:


\NewDocumentCommand\rqOne{}{Which origin methods impact CG sizes the most?}
\NewDocumentCommand\rqTwo{}{How local are derivatives of the most common origins?}
\NewDocumentCommand\rqThree{}{What are the effects of \tool{OriginPruner} on the size and usefulness of CGs?}
\NewDocumentCommand\rqFour{}{What is the computational overhead of \tool{OriginPruner}?}

\begin{description}
\item[\RQ{1}] \rqOne
\item[\RQ{2}] \rqTwo
\item[\RQ{3}] \rqThree
\item[\RQ{4}] \rqFour
\end{description}

By answering these research questions, we found several dominating origin methods, most of which are related to the Java type \code{Object} or the collection API.
We have implemented a vulnerability propagation analysis to investigate the effect of CG pruning on an exemplary client analysis.
We found that origin methods are mostly local and seem to be prunable without affecting our downstream analysis. 
Using \tool{OriginPruner} we can reduce the CG size by 14-58\%, which translates to a substantial boost in the analysis time of the vulnerability propagation, while only observing a marginal effect on the vulnerability results of vulnerability analysis.
Our results show that an ML-based approach is on par with a simple (non-ML) heuristic, which can be computed much faster, which is a promising direction for future research.

\smallskip
\noindent
Overall, this paper presents the following main contributions:

\begin{itemize}
\item We propose a novel CG pruning approach that leverages the method origin.
\item We show that the origin methods with the most significant effects on CG sizes are usually local and typically have low effects on inter-procedural analysis, making them a good candidate for pruning.
\item Through our evaluation of a vulnerability propagation analysis, we illustrate the need for better feature engineering for ML-based pruning models, as current state-of-the-art pruners cannot outperform a basic pruning heuristic.
\end{itemize}

The data and source code of this paper is available as an anonymous replication package~\cite{reppkg}. 

\section{Background}\label{sec:motive}
~\label{sec:motive}
Static call graphs build the foundation for many static software analyses.
They represent the calling relationships between methods within a program and model which \emph{source} methods call which \emph{target} methods.
Static analyses can easily extract the \emph{static call sites} from the abstract syntax tree of a program, which represents the locations in which an invocation is supposed to take place.
However, the static call site might refer to an interface type without implementation of that method, or during runtime, the target type is a more specific subtype of the static type.
For example, imagine a method with a parameter of type \code{Object}, on which \code{toString} is called, while likely thousands of \code{toString} implementations exist in all loaded classes out of which one will be the actual target of the call in a program run.
The challenge is to identify the set of implemented methods that might be called when the call site is dispatched.
To be sound, a call graph needs to be able to represent all possible executions.
To achieve this, CG generators usually perform a \emph{class-hierarchy analysis} to expand a local call site and include invocations of all known implementations (interface methods) or extensions (overridden methods) that exist throughout the code base and its dependencies.
Naturally, this includes very commonly overridden methods such as \code{Object.toString}.
Including edges for these methods ensures that the call graph accurately reflects all possible method invocations to make the static analysis sound, which is crucial for tasks like impact analysis, optimization, and vulnerability analysis.

\paragraph{Identifying Problematic Methods}
Especially methods that are defined in Java key classes, e.g., in the collection API, have an abundance of implementations in every code base and are used virtually everywhere (e.g., \code{Iterator.next}).
Including all these subtype invocations leads to a significant increase in the size and complexity of the static call graph, potentially making analyses less efficient and more difficult to interpret.

At the same time, most of these basic methods are only used to access data of data structures.
Our intuition says that these methods will be mostly local and that they would not affect static analyses like taint analyses or vulnerability propagation analyses, which follow inter-class data flow.
In this paper, we explore whether we can automatically identify such problematic methods and whether it is possible to guide a CG pruning by starting from only these problematic cases.

\paragraph{Origin Analysis}
While different implementations of the same interface or overridden versions of a method can have completely different method behavior, a good design should be constrained by the \emph{contract} that is established by the first definition of a method signature (i.e., \emph{design by contract}~\cite{meyer2002design}).
We refer to these first declarations as \emph{origin methods}.
As formulated in \emph{Liskov's Substitution Principle}~\cite{LSP}), these origin methods define requirements for subtyping relation.
The principle essentially states that an overridden method must be usable as a drop-in replacement by an unaware caller of the base method, for example, it should not introduce new and unanticipated side-effects or throw \code{Exception} types that clients of the base method do not expect.
In our processing, we will link every method in the dataset to its \emph{origin method}.

\paragraph{Pruning Call Graphs}
The task of CG pruning is initiated with a static CG denoted as $\mathbb{G}$, a directed graph that is generated via static analysis.
The nodes $V$ within this graph symbolize the defined methods, each distinguished by their method signature that includes the name, parameters, and return type.
The edges $E$ between these nodes are expressed as tuples and mark the invocation of one method (\emph{callee}) by another (\emph{caller}).
The CG pruner transforms the set of edges $E^\prime = f_{prune}(E)$ through a pruning function $f$, which uses heuristics or ML-based classifiers to decide on the edges that should be retained.
In this work, we will propose a novel tool that uses information about origin methods to guide the pruning decision.
A refined call graph, $\mathbb{G}^\prime= (V, E^\prime)$, is generated that preserves all vertices, but uses the transformed edge set.

\section{Related Work}
\paragraph{Call Graph Construction}
The study and development of call graphs have been a subject of significant interest. (ML-based) call graph pruning techniques, which do not incorporate real-time data, are categorized under static methods for call graph generation \cite{murphy1998empirical, reif2019judge, sui2020recall}. Conversely, dynamic analysis methods \cite{xie2002empirical, hejderup2018software} have been shown to reduce false positives and improve precision, albeit at the cost of scalability.

Additionally, efforts have been made to enhance the accuracy of call graphs. Lhotak \cite{lhotak2007comparing} introduced an interactive tool designed to elucidate the discrepancies between various static and dynamic analysis tools. Sawin and Rountev \cite{sawin2011assumption} proposed heuristics to better handle dynamic features such as reflection, dynamic class loading, and native method calls in Java, which improved the precision of the \emph{Class Hierarchy Analysis} (CHA) algorithm \cite{dean1995optimization}, while maintaining satisfactory recall.
Moreover, Zhang and Ryder \cite{zhang2007automatic} focused on producing accurate application-specific call graphs by identifying and eliminating false-positive edges between the standard library and the application itself. Recently, Antal et al.~\cite{antal2023javascript} studied the challenge of generating precise call graphs for JavaScript due to its dynamic nature, and hence they conducted a comparative study of static and dynamic CG generation tools for JavaScript. They found that while dynamic tools offer perfect precision, the recall of both static and dynamic approaches are very similar, ranging from 58\% to 69\%.
Keshani et al. proposed \tool{Frankenstein}~\cite{keshani2024frankenstein}, a fast and lightweight CG construction technique for software builds. The main idea is to create partial call graphs for each dependency in a program and then merge the resulting CGs into one whole program CG. Their approach is faster and has a small memory footprint, which makes it suitable for software build systems.

\paragraph{Machine learning for enhancing call graphs}
Currently, there are two notable ML-based models for call graph pruning: \tool{CGPruner}~\cite{utture2022striking} and \tool{AutoPruner}~\cite{le2022autopruner}. Utture et al. introduced \tool{CGPruner}, an ML-based approach aimed at reducing the false-positive rate of static analysis tools to enhance their appeal to developers \cite{utture2022striking}.
\tool{CGPruner} prunes unnecessary false-positive edges from the static call graph, which is integral to numerous static analyses. This is accomplished through a pre-execution learning process that involves the application of static and dynamic call-graph constructors, with dynamic call graphs utilized solely during the training phase on a set of training programs. This approach notably reduced the false-positive rate, in certain instances, from 73\% to 23\%.

However, \tool{CGPruner} does not analyze the semantics of the source code. To overcome this shortcoming, Le-Cong et al.~\cite{le2022autopruner} introduced \tool{AutoPruner}, which aims to eliminate false positives within call graphs by harnessing both the structural and semantic statistical information from the source code of caller and callee functions \cite{le2022autopruner}.
\tool{AutoPruner} employs \tool{CodeBERT} \cite{feng2020codebert}, a pre-trained Transformer-based model \cite{vaswani2017attention} designed explicitly for code. It fine-tunes this model to discern semantic features of each edge, integrates these with manually crafted structural features, and utilizes a neural classifier to classify each edge as either true or false positive.

A recent study of Mir et al.~\cite{mir2024effectiveness} compared the effectiveness of ML-based call pruning and established the benchmark dataset \tool{NYXCoprus} for this task.
Among their findings, they found that \tool{CGPruner} and \tool{AutoPruner} produce pruned call graphs with notoriously low recall, which renders these approaches unsuitable for security-focused applications like vulnerability analysis.
To alleviate this issue, they introduced a conservative strategy to fine-tune the code language models (CLMs) to maintain a high recall, close to static CGs, and benefit from better precision. This made pruned call graphs produced by CLMs useful for vulnerability analysis while benefiting from faster analysis.

So far, all the discussed papers have tackled the CG pruning problem with machine learning. There is also a recent work on discovering true edges for JavaScript's call graphs~\cite{bhuiyan2023call}. Their proposed technique combines structural and semantic information and employs Graph Neural Networks (GNNs)~\cite{scarselli2008graph} to identify true edges. The experimental results show a significant improvement to true positive rates in vulnerability detection.

\section{Methodology}

\begin{figure*}
    \centering
    \includegraphics[width=.9\textwidth]{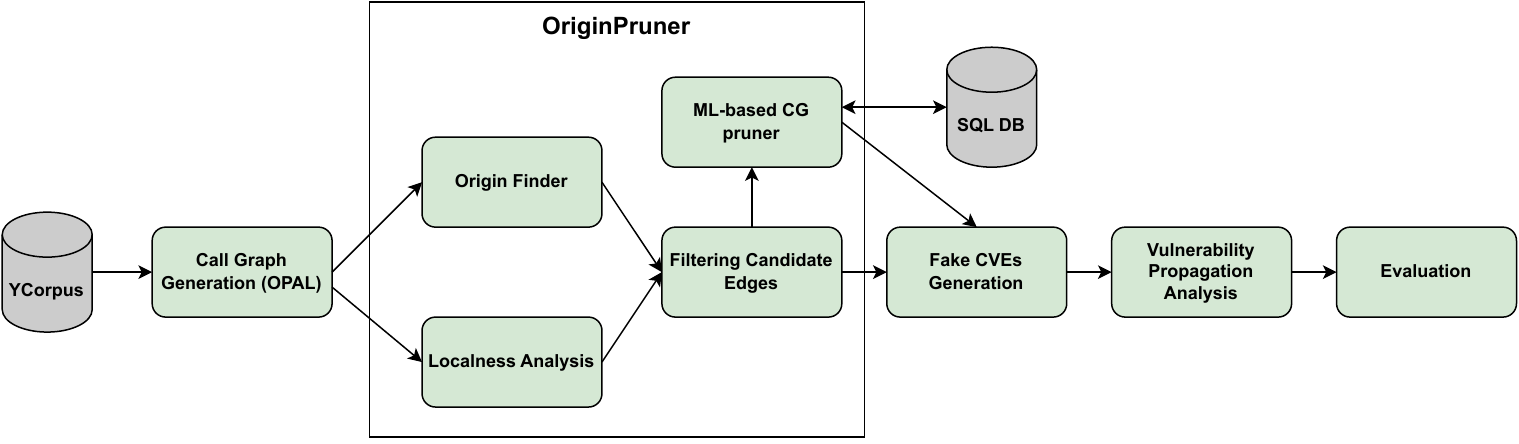}
    \caption{Overview of our proposed approach}
    \label{fig:overview}
\end{figure*}

Figure~\ref{fig:overview} shows the overview of our approach. We used Java programs from the \tool{YCorpus} dataset~\cite{mir2023effect} including their transitive dependencies to feed our pipeline. First, we generate static call graphs as a baseline for our evaluation.
Then, we perform two types of analyses on these call graphs, namely, origin finder and localness analysis. The origin finder identifies places, in which method signatures have been first introduced. The localness analysis then determines how local these methods are, i.e., if the methods rely solely on the Java standard library or use functionalities from methods implemented in the dependencies. Given the results of these two analyses, we identify a set of origin methods that we use as candidates for edge pruning in static CGs.
These candidates get pruned in two different strategies: a simply heuristic exhaustively prunes all calls that are related to these popular origin methods, a second strategy employs an ML-based CG pruning model to make the pruning decision.
We implement a vulnerability propagation analysis to evaluate the effect of the pruning on the reachability of vulnerabilities.
The following subsections provide more details about the individual steps.

\subsection{Call Graph Generation}\label{subsec:cg_gen}
To create static call graphs, we employed the OPAL framework~\cite{eichberg2014software, helm2020modular}, an advanced and well-established tool tailored for bytecode analysis. This methodology is based on employing a \emph{Class Hierarchy Analysis} (CHA), which represents a straightforward approach to identify call targets, by introducing edges to all subclasses of a type that override a given method.
Specifically, the OPAL framework generates the static call graphs from the Java bytecode, aiming to pinpoint potential method invocations to concrete methods. This process creates a directed graph, where methods are depicted as nodes, and the possible calls between them are represented as edges.
A crucial aspect of this analysis is to involve not just the application's bytecode but also this of its (transitive) dependencies and the Java Runtime Environment (JRE)'s core library \code{rt.jar} itself, which contains the standard Java API.
By analyzing \code{rt.jar}, the CHA can construct a comprehensive class hierarchy, which lays the foundation for resolving virtual method calls, a task achieved by identifying all potential callee methods in light of Java's polymorphism and inheritance mechanisms.

\subsection{Origin finding}\label{subsec:origin}
\label{sec:origin_finding}
As described in Section~\ref{sec:motive}, certain methods can cause an \emph{explosion} in the number of edges in call graphs. In this paper, we define them as \emph{origin methods}.
They play a pivotal role in understanding and managing the complexity of call graphs in object-oriented programming like Java. These are the methods where a particular method signature is first introduced in a class hierarchy, acting as the root for any subsequent overrides or derivative methods. The concept of origin methods is crucial because it helps identify the source of method signatures that proliferate across subclasses through inheritance and polymorphism. When a method in a superclass is overridden by multiple subclasses, each override is considered a derivative method. This inheritance and overriding mechanism can lead to a significant increase in the number of edges in a call graph, especially for origin methods high up in the type hierarchy with many subclasses. The explosion of edges attributable to these origin methods reflects the potential paths of method invocations, which can be particularly challenging to analyze due to the increased complexity and size of the call graph.

To tackle the challenge posed by origin methods and their derivatives in static analysis, we proposed an algorithm that is designed to identify and assess the impact of these methods on the structure of call graphs. This algorithm is outlined in the provided pseudocode in Algorithm~\ref{alg:origin}.

\begin{algorithm}[b]
    \caption{Finding origin methods of a call graph}
    \label{alg:origin}
    \input{algos/finding-origin}
\end{algorithm}

The algorithm systematically examines each edge in a call graph to find the origin method responsible for the method invocation represented by the edge. It iterates through the edges of the CG, tracing each target method's type back through its hierarchy until it finds the highest type where the method signature was first declared. By mapping each method to its origin, this approach allows us to investigate the extent to which a single origin method can influence the call graph, quantifying the number of edges it effectively generates. This information is invaluable for optimizing CG analyses, as it highlights the methods that are most influential in the complexity of the graph, guiding efforts to simplify or focus the analysis on the most impactful areas.

The identification of \emph{origin methods} through the aforementioned algorithm enables a systematic way of investigating the impact of inheritance in CGs.
The intuition is that a small number of origin methods cause large parts of the CG.
Pruning these methods would systematically reduce the CG size and, in extensions, significantly enhance the efficiency and focus of downstream static analysis.
The approach is particularly useful because, in contrast to ML-based approaches, it allows to preserves the integrity of the analysis by maintaining the critical paths of method invocations by reducing CG complexity without substantial loss of essential information that goes beyond the identified local methods.
This enhances the practicality of static analysis tools, making them more adaptable to large-scale software projects where unpruned CGs would otherwise be impractical to work with.

\subsection{Identifying Localness Levels}\label{subsec:localness}
The concept of localness in relation to the derivatives of an origin method introduces a nuanced framework for analyzing the scope and impact of method invocations within an application. This framework categorizes methods based on the extent of their interactions with other components of the system, ranging from purely internal functionality to interactions that cross package boundaries.
We define four categories of localness:

\begin{itemize}
    \item \emph{Level 0}: A method does not call anything or it only calls Java functionalities.
    \item \emph{Level 1}: A method does call other functionalities than Java but it does not exit its class or its class hierarchy.
    \item \emph{Level 2}: A method calls at least one function from outside of its class hierarchy, but it remains within the same project.
    \item \emph{Level 3}: A method calls at least one function from outside of its class hierarchy and the target of this call is also in another package.
\end{itemize}

Labeling the target methods within a CG with these localness levels helps to better understand an application's architecture and the dependencies between its components.
For instance, a derivative method with a localness level of 0 or 1 indicates a high degree of cohesion within the class or class hierarchy, suggesting that the functionality is tightly integrated and likely uninteresting for data-flow analyses.
On the other hand, a derivative with a localness level of 3 points to a broader scope of interaction, which might be necessary for the application's functionality but also introduces more dependencies, which increases the complexity and the likelihood to be relevant for complex data-flow analyses.
Algorithm \ref{alg:categorize} shows how to find the localness level of target methods in call graphs.

\begin{algorithm}[t]
    \caption{Localness algorithm}
    \label{alg:categorize}
    \input{algos/localness}
\end{algorithm}

\subsection{Pruning Strategy}
After performing the origin-finding analysis, we have the origin methods and their derivatives for Java programs. These methods are frequently overridden or extended across different classes and packages within programs in the dataset. By considering Top-$n$ most frequent origin methods, we create an "exclusion list" of $n$ elements. Given this, we prune call graphs edges for which their target type is a derivative of an origin method in the list. Algorithm~\ref{alg:prune_cg} shows pseudo-code for pruning call graph edges using the described pruning strategy.

\begin{algorithm}[t]
    \caption{Pruning call graphs with a pruning strategy}
    \label{alg:prune_cg}
    \input{algos/pruning}
\end{algorithm}

\subsection{Dataset}
For this work, we used the \tool{YCorpus} dataset~\cite{mir2024effectiveness}. It contains 23 Java projects that all have more than 80\% test coverage. Its programs contain the JARs of transitive dependencies plus source code. Despite its small size, YCorpus is suitable for evaluating (ML-based) call-pruning models, as it contains real-world Java projects such as Apache Commons IO, AssertJ, and MyBatis 3, and each project has an average of around 50K lines of code.

\subsection{ML-based call graph pruning}
We want to train an ML model as a baseline to compare our origin-based pruning to.
ML-based call graph pruning aims to prune superfluous or false edges in static call graphs by learning from actual program execution paths, i.e., dynamic call graphs.
For this work, we employ the well-known code language model \tool{CodeBERT}~\cite{feng2020codebert}, which leverages semantics embedded within code.
We follow previous work and uses a \emph{paranoid pruning} with weight of 0.95 in the learning process and a confidence level of over 95\%~\cite{mir2023effect}.
This method prioritizes the retention of edges, i.e., very cautiously pruning only when there is high confidence, thus minimizing the risk of losing true call graph edges. With the described strategy, CodeBERT not only enhances the precision of call graphs but also maintains high recall, almost identical to that of static call graphs, which is crucial for downstream analyses, such as vulnerability detection at the method level.

We employ the ML-based CG pruner in combination with our proposed approach, which leverages the results of the origin finder and localness analysis to prune edges. Specifically, we query the ML model to decide whether a candidate edge for pruning, i.e., a 
derivative of the origin methods, should be pruned. We expect that this makes our approach more conservative. In the paper, we refer to this combination as \tool{OriginPruner} with Selective, ML-based pruning.

\subsection{Generating Artificial Vulnerabilities}
To study the effect of our proposed call graph pruning on vulnerability propagation, we follow the methodology of previous work and create artificial CVEs to make the evaluation more scalable~\cite{mir2023effect}.
The focus of our evaluation is not a representative selection of vulnerable methods, we only need a high number of vulnerabilities that allows us to instantiate our reachability-based evaluation with different configurations.

We first separate application nodes from those of dependencies. This distinction is crucial since the aim is to inject vulnerabilities exclusively into the dependency nodes. We randomly mark 100 of these dependency nodes to be associated with one of the artificial vulnerabilities. The selection should be random to ensure that the simulated vulnerabilities are spread unpredictably across the dependencies, mirroring the nature of real-world vulnerabilities.

\subsection{Vulnerability Propagation Analysis}\label{subsec:vpa}
To find whether vulnerable nodes/methods in static call graphs are reachable from the main application's methods, we perform an inverse Breadth-First Search (BFS) from nodes identified as vulnerable in the project's dependencies.
We already know the vulnerable nodes in the CG and hence we initiate the search from these nodes rather than from all applications nodes.
Through the backwards traversal, we can pinpoint the sequences of calls that exposes the application to risks emanating from its dependencies.
This is crucial for understanding the impact of vulnerabilities in third-party libraries and enabling developers to prioritize and remediate threats more effectively.

\subsection{Implementation}
We implemented most of the components in our proposed approach in Java 11 such as CG generation, origin finding, localness categorization, vulnerability analysis, and artificial CVEs generation. We used \tool{Apache Kafka} to stream data into our end-to-end pipeline. We employed the \tool{JGraphT} framework~\cite{jgrapht} to work with graphs and implement vulnerability propagation analysis. Also, we implemented an ML-based CG pruner, a fine-tuned \tool{CodeBERT}, as an inference service in Python 3.10 using \tool{PyTorch} 2.2~\cite{pt2,paszke2019pytorch} and \tool{Ray Serve}~\cite{moritz2018ray}. We used the \tool{PostgreSQL} database to cache the predictions of the inference service. This is helpful to speed up the ML-based CG pruning process when we consider different sizes of filters in our experiments.

We conducted all the experiments in this paper on a workstation machine with Ubuntu 22.04 LTS, 2xAMD EPYC 7H12 64-core processor, and 512 GB of RAM. The ML-based CG pruner model inference was done on an RTX 3080 10 GB.

\section{Results}
\subsection{RQ1: \rqOne}
We study the origin methods in call graphs as they can provide insights to better understand the complexity of Java applications. Fundamentally, origin methods introduce a method signature subsequently overridden by derivative methods across various types. In other words, the proliferation of derivative methods from a single origin can dramatically increase the complexity of call graphs, which map the interactions and dependencies between different parts of a software system. This complexity is not merely a research problem but also has practical implications for software maintenance, optimization, and vulnerability analysis. By studying the impact of origin methods on the expansion of call graphs, we can identify patterns and anomalies that help us to form a set of heuristics or rules for pruning call graph edges.

\paragraph{Methodology}
To find origin methods in the YCorpus dataset, we first generate static call graphs using OPAL, which is explained in Section~\ref{subsec:cg_gen}. By iterating through the edges of the generated static CGs, we follow the origin-finding procedure to find the origin of target nodes. We aggregate each project's discovered origin methods to report the Top-10 origin methods in the dataset. In addition, we report the number of unique derivatives for the Top-10 origin methods.

\paragraph{Results}
Table~\ref{tab:origin_freq} shows Top-10 most frequent origin methods for YCorpus. \code{Iterator.next()} is the most common origin method with over 15M frequency, which is used to iterate over elements in data structures like \code{Set}, \code{List}, etc. Considering this, our intuition in Section~\ref{sec:motive} is backed up, providing that methods like \code{toString()} or \code{next()} are very frequent in static call graphs. Moreover, it can be observed that \code{XWrapperBase.toString()} is the 8th most common origin method, which is part of Java Runtime Environment (JRE) and is used for building graphical user interfaces, particularly for X11 windowing systems. As described in Section~\ref{subsec:cg_gen}, we included JRE libraries when building static call graphs.

Also, Table~\ref{tab:origin_der} shows the number of unique derivatives for the Top-10 frequent origin methods. Interestingly, \code{Iterator.next()} does not have the most unique derivates given that it is the most common origin method in the YCorpus dataset. In fact, \code{PrivilegedAction.run()} has the most unique derivatives among other origin methods. Overriding this method allows developers to encapsulate security-sensitive operations in a single, reusable component. This can make the code more readable and maintainable, as the security-related code is localized and not scattered throughout the application. Overall, considering the most frequent origin methods and their derivatives, our observation here hints at the fact that these methods can potentially be a candidate for pruning edges in static call graphs. We should also point out that we only reported the Top-10 most common origin methods due to the limited space in the paper. Readers can refer to our replication package for the extensive list.

\begin{table}[!t]
\centering
\caption{Top 10 most common origin methods in YCorpus}
\label{tab:origin_freq}
\begin{tabular}{@{}lr@{}}
\toprule
Method\tnote{a} & Frequency \\
\midrule
/j.u/Iterator.next()/j.l/Object & 15,795,483 \\
/j.u/Iterator.hasNext()/j.l/BooleanType & 13,691,858 \\
/j.l/Iterable.iterator()/j.u/Iterator & 5,243,680 \\
/j.u/Collection.size()/j.l/IntegerType & 2,317,670 \\
/j.u/Map.get(/j.l/Object)/j.l/Object & 2,274,769 \\
/j.u/Map.put(/j.l/Object,/j.l/Object)/j.l/Object & 1,364,892 \\
/j.u/List.get(/j.l/IntegerType)/j.l/Object & 1,211,656 \\
/s.a.X11/XWrapperBase.toString()/j.l/String & 869,519 \\
/j.u/Collection.contains(/j.l/Object)/j.l/BooleanType & 693,782 \\
/j.u/Enumeration.nextElement()/j.l/Object & 693,372 \\
\bottomrule
\end{tabular}

\vspace{4pt}
\emph{Note:} For brevity, package names are \\ abbreviated with their first letter. 

\vspace{6pt}
\caption{No. of unique derivatives for the most common origin methods}
\label{tab:origin_der}
\begin{tabular}{@{}lr@{}}
\toprule
Origin methods & \#Derivatives \\
\midrule
/j.s/PrivilegedAction.run()/j.l/Object & 18,465 \\
/j.u/ListResourceBundle.getContents()/j.l/Object[][] & 7,576 \\
/j.i/Closeable.close()/j.l/VoidType & 5,945 \\
/j.u/Iterator.next()/j.l/Object & 5,370 \\
/j.l/Iterable.iterator()/j.u/Iterator & 4,842 \\
/j.u/Iterator.hasNext()/j.l/BooleanType & 4,692 \\
/j.u/Collection.size()/j.l/IntegerType & 4,644 \\
/j.u/Iterator.remove()/j.l/VoidType & 4,282 \\
/j.s/PrivilegedExceptionAction.run()/j.l/Object & 3,987 \\
/j.u.f/Function.apply(/j.l/Object)/j.l/Object & 3,691 \\
\bottomrule
\end{tabular}
\end{table}


\begin{figure}
    \centering
    \includegraphics[width=\columnwidth]{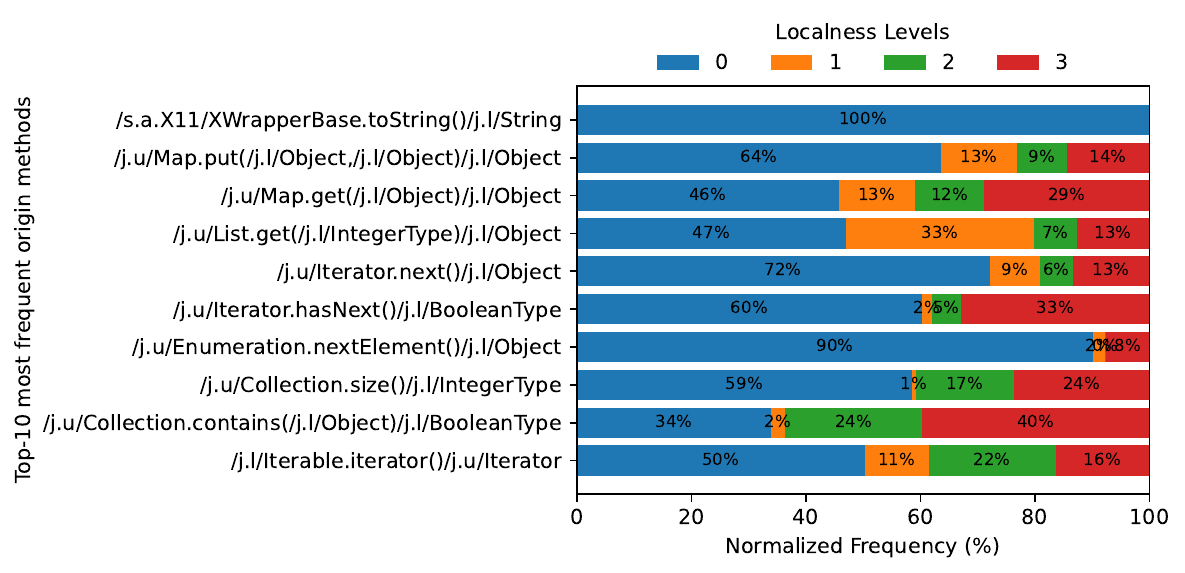}
    \vspace{-20pt}
    \caption{Relative frequency of localness levels for the derivatives of the top 10 origin methods}
    \label{fig:localness}
\end{figure}


\subsection{RQ2: \rqTwo}
We investigate the localness of origin methods as it can provide additional valuable insight for pruning call graphs to reduce their size and complexity. By categorizing origin methods and their derivatives based on their scope of interaction, we can pinpoint critical junctures where vulnerabilities are likely to propagate or unnecessary paths could be eliminated through a systematic pruning of call graphs.

\paragraph{Methodology}
Given the generated CGs for YCorpus from RQ1, we iterate over the nodes to determine how local they are by following the algorithm for the localness analysis, explained in Section~\ref{subsec:localness}. Next, we label the derivatives of the origin methods, found in RQ1, from Level 0 to Level 3. We aggregate the number of each level for all the derivatives of the Top-10 origin methods and report the relative frequency of the four labels for the Top-10 origin methods.

\paragraph{Results}
Figure~\ref{fig:localness} shows the relative frequency of localness levels for the derivatives of the Top-10 origin methods. First, we observe that 72\% of derivatives of the most common origin method \code{Iterator.next()} has Level 0 localness. This means that they do not call any other methods or may call only internal JRE functions. In other words, the derivatives of the method \code{Iterator.next()} are quite "local", in the sense that they can be removed from call graphs without potentially losing crucial paths or information. Overall, it can be seen that most of the methods that override one of the Top-10 origin methods are quite local. They call either no other methods or call method(s) from their inherited class hierarchy. Of these 10 origin methods, interestingly, only \code{Collection.contains(Object)} has around 63\% of derivatives that call method(s) outside its class hierarchy or program's dependencies. One more observation is all the derivatives of the origin method \code{XWrapperBase.toString()} have a localness of Level 0. This is expected as this method is part of the JRE library and does not use any external functionalities. Given the results of this RQ, we are now more confident in using a pruning strategy based on the common origin methods as their derivatives are mostly Level 0 or 1 in terms of localness.

\subsection{RQ3: \rqThree}
In the previous two RQs, we studied the Top-10 most common origin methods in call graphs and how local their derivatives are. Using the gained insights and findings, we prune edges in call graphs to study the effect of pruning on the size and complexity of call graphs. Also, we use the vulnerability propagation analysis as a case study to investigate the effectiveness of our proposed pruning technique for security applications where soundness is essential. In other words, if one vulnerable method is missed in the analysis due to pruning, it can be costly for developers or users by allowing attackers to exploit the vulnerable method. 

\paragraph{Methodology}
First, we prune edges call graphs by following our filtering strategy, which is an exclusion list of Top-1000 frequent origin methods. Specifically, we consider a filter of different sizes from the set $\{1,2,3,5,10,25,50,100,1000\}$ to prune edges. Basically, for each program in YCorpus, we end up with nine different pruned call graphs. Given 203 pruned CGs, we report the average and standard deviation for nodes, edges, and relative size reduction for every nine different filters, compared to the baseline, which is OPAL's static CGs. In addition, we evaluate our proposed approach in combination with an ML-based CG pruning technique, called \tool{OriginPruner} with Selective, ML-based Pruning. This approach gives edges to derivatives of the origin methods to the ML model which makes the final decision for pruning.

Second, we perform a vulnerability propagation analysis (described in Section~\ref{subsec:vpa}) for OPAL's static CG and all the pruned CGs from the previous step, given the artificial CVEs. We report the average number of vulnerable paths, reachable vulnerable nodes, and analysis time for OPAL's static cg as a baseline and every nine different filters or an exclusion list. We also repeated this experiment for  \tool{OriginPruner} with Selective, ML-based pruning. To accurately measure analysis, we consider JVM's Just-in-Time compilation by warming up the analysis program, running it three times, and reporting the average analysis time across runs. 

\NewDocumentCommand\OP{m}{OriginPruner$_{#1}$}

\paragraph{Results}
Table~\ref{tab:cg_size} shows the effect of pruning call graph size.
We compare the baseline results of an OPAL call graph with different configurations of \OP{N}, which prunes calls to all methods that are related to the Top-N origin methods, either \emph{exhaustively} or \emph{selectively} using an ML-based pruning classifier.
OPAL's static CGs are shown as a baseline at the top of the table. On average, across all programs in YCorpus, there are around 186K and 4.9M nodes and edges, respectively.
Considering exhaustive pruning, just pruning the Top-1 origin method \code{Iterator.next()} already reduces the size of call graphs from 4.9M to 4.22M on average, a reduction in size by 14\%. Notably, pruning the Top-10 origin methods reduces the CG size by 37\% to 3M edges on average.
This implies that, on average, more than one-third of edges are a call to a method that overrides one of those Top-10 most common origin methods reported in Table~\ref{tab:origin_freq}.
From Top-10 to Top-1000, we still observe a steady decrease in the size of call graphs to the point where Top-1000 reduces the number of edges by more than half on average, i.e., 58\%.

Compared to exhaustive pruning, selective pruning uses an ML-based classifier to make individual pruning decisions for each edge.
This approach also consistently reduces the size of call graphs, even though it prunes fewer edges. For instance, selectively pruning the Top-10 origin methods has almost the same effect on the CG size as exhaustive pruning of the Top-5 origin methods.
It is interesting to see that the ML-based CG pruner seems to implicitly learn the concept of origin methods from its training data: the underlying CodeBERT model is usually very conservative in its pruning decisions, yet it learns that a less aggressive pruning is required for the origin methods.

Table~\ref{tab:vuln_analysis_results} shows the effect of call graph pruning on the vulnerability propagation analysis. Starting with OPAL as a baseline, we observe that, on average, there are around 18K reachable paths and 76\% of 100 artificial CVEs are reachable from application nodes.
The artificial CVEs are randomly distributed among CG nodes, which means that they are also not necessarily reachable in the unpruned static call graphs.
When pruning the Top-1 origin methods, \tool{OriginPruner} improves the analysis time by around 19\% while preserving the same vulnerable node reachability through vulnerable paths as the unpruned CG (i.e., 76\%).
This trend continues when more is pruned.
By considering the derivatives of Top-10 most common origin methods for pruning, we observe a small 2\% decrease in the average number of reachable vulnerable nodes while the analysis time can be reduced from 72s to 49s (33\% faster).
Using \tool{OriginPruner}, it is, therefore, possible to select a trade-off between CG accuracy and size by selecting the number of origin methods to prune. Accepting a small decrease in the soundness of the vulnerability analysis can result in a significant lowering of the computational time.
We also see that there is a sweet spot.
Pruning Top-100+ origin methods substantially reduce the CG sizes, but the effects on reachability become larger with a 10+\% decrease in reachable vulnerable nodes. This makes these configurations impractical for vulnerability analysis as it can cause developers to miss the presence of vulnerabilities in their application or its dependencies.
With the selection, it is possible to obtain the same number of reachable vulnerable nodes as OPAL while using a Top-3 filter for pruning. Overall, the results of the selective approach show that the ML-based CG pruner makes our proposed approach slightly more conservative by pruning fewer edges. However, the practicality of the selective approach is still a question given its additional computational overhead.

\begin{table*}
\centering
\caption{The effect of pruning on the size of call graphs and vulnerability propagation}
\label{tab:cg_size}
\label{tab:vuln_analysis_results}
\begin{tabular}{@{}lccccrr@{}}
\toprule
&&&& \multicolumn{2}{c}{{\bf Reachable ...}} & \\
\cmidrule{5-6}
\textbf{Approach} & \textbf{\#CG Nodes} & \textbf{\#CG Edges} & \textbf{Reduction (\%)} & {\# Paths} & {Nodes (\%)} & \textbf{Analysis Time (s)} \\
\midrule
OPAL & 186k $\pm$ 27K & 4.9M $\pm$ 1.5M & 0.0 & 18.8K $\pm$ 42.8K & 0.76 $\pm$ 0.17 & 72.94 $\pm$ 25.47 \\
\midrule
\multicolumn{7}{c}{\tool{OriginPruner} with Exhaustive Pruning} \\
\midrule
\OP{1} & 186K $\pm$ 27K & 4.2M $\pm$ 1.3M &  0.14 $\pm$ 0.01 & 18.7K $\pm$ 42.3K & 0.76 $\pm$ 0.17 & 61.22 $\pm$ 18.18 \\
\OP{2} & 186K $\pm$ 27K & 3.6M $\pm$ 1.1M &  0.26 $\pm$ 0.02 & 18.6K $\pm$ 41.8K & 0.75 $\pm$ 0.17 & 57.55 $\pm$ 15.84 \\
\OP{3} & 186K $\pm$ 27K & 3.4M $\pm$ 1.0M &  0.29 $\pm$ 0.03 & 18.5K $\pm$ 41.8K & 0.75 $\pm$ 0.17 & 55.71 $\pm$ 15.83 \\
\OP{5} & 186K $\pm$ 27K & 3.3M $\pm$ 0.9M &  0.33 $\pm$ 0.04 & 18.5K $\pm$ 41.8K & 0.75 $\pm$ 0.17 & 53.49 $\pm$ 15.27 \\
\OP{10} & 186K $\pm$ 27K & 3.0M $\pm$ 0.9M &  0.37 $\pm$ 0.04 & 18.2K $\pm$ 41.6K & 0.74 $\pm$ 0.17 & 49.59 $\pm$ 12.87 \\
\OP{25} & 186K $\pm$ 27K & 2.8M $\pm$ 0.7M &  0.43 $\pm$ 0.05 & 18.1K $\pm$ 41.6K & 0.73 $\pm$ 0.16 & 47.76 $\pm$ 12.23 \\
\OP{50} & 185K $\pm$ 27K & 2.6M $\pm$ 0.7M &  0.47 $\pm$ 0.05 & 17.7K $\pm$ 40.5K & 0.72 $\pm$ 0.16 & 43.83 $\pm$ 10.41 \\
\OP{100} & 185K $\pm$ 27K & 2.4M $\pm$ 0.6M &  0.50 $\pm$ 0.05 & 15.9K $\pm$ 37.2K & 0.65 $\pm$ 0.15 & 38.82 $\pm$ 9.15 \\
\OP{1000} & 180K $\pm$ 27K & 2.0M $\pm$ 0.5M &  0.58 $\pm$ 0.05 & 13.0K $\pm$ 30.6K & 0.55 $\pm$ 0.13 & 29.76 $\pm$ 6.90 \\
\midrule \multicolumn{7}{c}{\tool{OriginPruner} with Selective, ML-based Pruning} \\ \midrule
\OP{1} & 186K $\pm$ 27K & 4.3M $\pm$ 1.4M &  0.13 $\pm$ 0.01 & 18.7K $\pm$ 42.3K & 0.76 $\pm$ 0.17 & 57.12 $\pm$ 16.67 \\
\OP{2} & 186K $\pm$ 27K & 3.7M $\pm$ 1.2M &  0.25 $\pm$ 0.01 & 18.6K $\pm$ 41,8K & 0.76 $\pm$ 0.17 & 54.53 $\pm$ 16.57 \\
\OP{3} & 186K $\pm$ 27K & 3.6M $\pm$ 1.1M &  0.27 $\pm$ 0.02 & 18.6K $\pm$ 41.8K & 0.76 $\pm$ 0.17 & 54.22 $\pm$ 15.11 \\
\OP{5} & 186K $\pm$ 27K & 3.4M $\pm$ 1.1M &  0.31 $\pm$ 0.02 & 18.6K $\pm$ 41.8K & 0.75 $\pm$ 0.17 & 51.23 $\pm$ 14.06 \\
\OP{10} & 186K $\pm$ 27K & 3.2M $\pm$ 1.1M &  0.34 $\pm$ 0.03 & 18.4K $\pm$ 41.7K & 0.75 $\pm$ 0.17 & 50.11 $\pm$ 14.12 \\
\OP{25} & 186K $\pm$ 27K & 3.0M $\pm$ 1.0M &  0.40 $\pm$ 0.03 & 18.3K $\pm$ 41.6K & 0.74 $\pm$ 0.17 & 48.35 $\pm$ 13.84 \\
\OP{50} & 185K $\pm$ 27K & 2.8M $\pm$ 0.9M &  0.44 $\pm$ 0.03 & 18.0K $\pm$ 40.6K & 0.73 $\pm$ 0.17 & 45.21 $\pm$ 12.65 \\
\OP{100} & 185K $\pm$ 27K & 2.6M $\pm$ 0.9M &  0.47 $\pm$ 0.03 & 16.5K $\pm$ 37.7K & 0.67 $\pm$ 0.16 & 41.55 $\pm$ 15.25 \\
\OP{1000} & 181K $\pm$ 27K & 2.2M $\pm$ 0.8M &  0.54 $\pm$ 0.03 & 13.9K $\pm$ 32.0K & 0.58 $\pm$ 0.14 & 33.16 $\pm$ 12.23 \\
\bottomrule
\end{tabular}
\end{table*}

\subsection{RQ4: \rqFour}
The previous RQ has shown that \tool{OriginPruner} can leverage origin methods to guide the CG pruning, which can greatly reduce the size of call graphs with minimal loss in the accuracy of the analysis.
While this speeds up the graph search, these benefits must need to be computed first as call-graph pruning is a post-processing step that adds additional overhead to the call-graph generation process itself.
Therefore, we want to investigate the additional computational cost of \tool{OriginPruner} compared to basic CG generation.

\paragraph{Methodology}
First, we report the average time of call graph generation using the OPAL framework for all the programs in YCorpus. Then, we report the average pruning time of \tool{OriginPruner} for processing the Top1-1000 origin methods.
Note that we do not report the computational cost of the origin finder and localness analysis as they only need to be executed once for a dataset, similar to training an ML-based CG pruner.
Lastly, we report the average time for \tool{OriginPruner} with Selective, ML-based pruning which involves querying a CodeBERT-based model for pruning edges found by our approach.

\paragraph{Results}
Table~\ref{tab:cg_pruning_time} shows the computational time of call graph generation and the computation time of \tool{OriginPruner}, both in the mutually exclusive, our proposed CG pruning approach, and the selective approach. The average time for CG generation using OPAL is around 30.5s for all programs in the dataset.
Also, the average time of the origin finder and localness analysis is about 100s. This is a one-time cost, meaning that it is similar to training an ML model, which needs to be done once for a dataset. We observe that exhaustive pruning only adds $\sim$9-10s ($\sim$33\%) overhead to the CG generation. For instance, to prune the Top-10 origin methods, the CPU time is around 9.2 seconds on average.
Such a small overhead can usually be justified by further time savings that a reduced CG size will introduce in other downstream analyses and a smaller memory footprint.

We also see that selective pruning is much more computationally expensive.
This is expected as the CodeBERT-based CG pruner model is queried for all the edges identified by \tool{OriginPruner}.
Even though we have provided caching mechanisms to speed up the model predictions, it becomes clear that the ML model does not scale well with relatively large CGs.

\begin{table}
\centering
\caption{Run-time of CG generation and pruning}
\label{tab:cg_pruning_time}
\begin{tabular}{@{}lrr@{}}
\toprule
& \multicolumn{2}{c}{{\bf Pruning Time [s]}} \\
\cmidrule{2-3}
\textbf{Approach} & Exhaustive & Selective \\
\midrule
Top-1 & 10.6 $\pm$ 3.1 & 726.9 $\pm$ 935.5 \\
Top-2 & 10.1 $\pm$ 3.0 & 705.1 $\pm$ 897.3 \\
Top-3 & 9.8 $\pm$ 2.9 & 694.4 $\pm$ 844.9 \\
Top-5 & 9.5 $\pm$ 2.8 & 611.4 $\pm$ 766.6 \\
Top-10 & 9.3 $\pm$ 2.7 & 545.7 $\pm$ 711.5 \\
Top-25 & 9.1 $\pm$ 2.6 & 521.8 $\pm$ 716.5 \\
Top-50 & 8.9 $\pm$ 2.6 & 481.8 $\pm$ 687.6 \\
Top-100 & 8.8 $\pm$ 2.5 & 439.2 $\pm$ 655.8 \\
Top-1000 & 8.8 $\pm$ 2.6 & 436.4 $\pm$ 639.0 \\
\bottomrule
\end{tabular}

\vspace{5pt}
\emph{Note:} All approaches require the same CG \\
generation (30.58s $\pm$ 5.06s) and Origin \\
and Localness analysis (99.96s $\pm$ 26.36s).
\end{table}

\section{Discussion}
In this section, we discuss the implications of the obtained results and directions for future research.

\paragraph{Promising Results}
Overall, our empirical results have shown that using an origin analysis to identify the locations in the call graph that are the main culprits for large CG sizes allows to better guide the pruning, so it only has marginal effects on analysis results.
It also seems feasible to make CG pruning part of the CG generation process. As shown in RQ4, CG pruning can be considered as a post-processing step, which adds a small overhead to CG generation, and it can even be configured to prune more or fewer origin methods, or disable the pruning altogether.
It also becomes clear that CG pruning can benefit from using truly semantic features over purely structural features even though it requires domain knowledge to tap this potential.
Future work should investigate whether other statically available facts, e.g., the distance of packages, or the existence of specific types in a class context, could be used to improve the exhaustive pruning heuristic.

\paragraph{Localness Levels}
In the RQ2, we have validated our intuition that most of the common origin methods are indeed very local to justify the exhaustive pruning.
However, the extracted localness level could also be used as a feature that can inform the pruning decision.
Maybe instead of just pruning all the Top-N origin methods, also the typical localness of subtypes for each of the Top-N origin method should be taken into consideration and \tool{OriginPruner} should only prune such methods from the Top-N that have a very big fraction of level 0 methods.
Future work could further differentiate our localness levels in the pruning or could devise new levels or an alternate definition of localness that might be better suited for the task.

\paragraph{Limited ML Performance}
CLMs have achieved a big success in software engineering tasks like code generation, thanks to their billion-parameter-scale size~\cite{hou2024large}.
For the CG pruning task, however, it looks like are currently not yet able to outperform simple heuristics.
It is interesting to see that an ML approach would pick up a strategy for pruning origin methods that is similar to our heuristic solution and, therefore, results in a similar performance.
It does so at a massive computational cost though, which is a strong limitation.
We think that future work is necessary in two dimensions to make ML-based pruning relevant.
First, it does not seem to be sufficient to train models with basic structural features or plain source code, as the semantic features are hard to pick up during model training.
Future work needs to investigate better feature engineering or code representation to make it possible to leverage semantic features that usually require a static analysis for extraction.
Second, it is necessary to improve the scalability of ML-based pruners to make them feasible in practice.
The most promising direction is to avoid treating the pruning as a post-processing step and instead make it an integral part of a hybrid CG generator.
Note though that this optimization is also available for heuristic approaches, which makes it a particularly interesting direction to explore.


\section{Threats to Validity}
The empirical results of our work are subject to several potential threats to validity.
We will introduce these threats and our mitigation strategies.

\paragraph{Correctness}
Our codebase is small and only has around 4K LoC (Java/Python).
The authors have reviewed the algorithms multiple times to prevent bugs.
Also, the CG generation and pruning have been integrated into the ABCDEF\footnote{Omitted for the double-blind review process.} research project and have been used and tested by other users.

\paragraph{Representativeness}
Our choice of ML model could negatively affect selective pruning.
To mitigate this problem, we have re-used a fine-tuned CodeBERT model that has been shown to be effective for CG pruning before~\cite{mir2023effect}.

For vulnerability propagation analysis, we followed the methodology of the previous work and randomly injected artificial CVEs to call graph nodes to assess the effectiveness of \tool{OriginPruner}~\cite{mir2024effectiveness}.
While we have not tested the evaluation with actual CVEs, we do not see any reason to doubt the representative of the artificial CVEs.
However, future work should investigate this assumption.

\paragraph{Dataset}
We used an existing dataset of 23 Maven libraries, YCorpus.
The libraries are representative, so we do not see any reason to believe that a larger dataset would exhibit different results in terms of origin analysis or localness.
As such, we believe that our findings also hold in larger datasets.
If at all, the reported numbers for origin methods will only increase in our favor for larger projects, as more subtypes exist, which emphasizes the problem even more.

\section{Conclusion}
In this paper, we presented \tool{OriginPruner}, a novel approach to call graph pruning, based on the origin methods and their locality to address the limitations in ML-based call graph pruning techniques, which are lack of generalization beyond training set and high inference cost. Our proposed approach leverages the insights from the origin methods, specifically, methods that introduce a signature in a class hierarchy and are frequently overridden, and their locality to prune unnecessary edges in call graphs efficiently. This approach significantly reduces the call graph size and complexity, making downstream analyses more practical for large-scale software projects. Based on the YCorpus dataset, our empirical findings reveal that specific origin methods, such as \code{Iterator.next()}, play a pivotal role in call graph complexity by being frequently overridden across different classes. We found that these methods and their derivatives have predominantly localness of Level 0 or 1, providing a solid basis for pruning without sacrificing the soundness required for tasks like vulnerability analysis. Moreover, the obtained results show that our proposed call graph pruning approach could reduce call graph size by up to 58\%, significantly improving analysis speed (up to 2.4x times) with no or minimal impact on the accuracy of downstream applications like vulnerability propagation analysis. Additionally, our approach was shown to be computationally more efficient than existing ML-based approaches, highlighting its practicality for real-world applications.

\balance
\bibliographystyle{IEEEtran}
\bibliography{reference}
\end{document}

%% file: cops.tex
\usepackage[utf8]{inputenc}
\usepackage{xparse} 

\usepackage{balance} 
\usepackage{cleveref}

\usepackage{booktabs} 

\usepackage{enumitem}
\usepackage{noindentafter}
\setlist{
	leftmargin=12pt,
	itemsep=2pt plus 1pt,
	topsep=2pt plus 1pt,
	parsep=2pt plus 1pt,
	listparindent=0pt,
}
\setlist[description]{
	font=\normalfont\itshape
}
\NoIndentAfterEnv{itemize}
\NoIndentAfterEnv{description}
\NoIndentAfterEnv{enumerate}

\usepackage[noindentafter]{titlesec}
\titlespacing{\paragraph}{0pt}{\medskipamount}{4pt}
\titleformat{\paragraph}[runin]{\itshape}{\noindent}{0pt}{}[.] 

\NewDocumentCommand\RQ{+m}{RQ$_{\normalfont #1}$}
\NewDocumentCommand\citeTodo{+m}{{\color{red}[?]}}

\NewDocumentCommand\code{m}{{\small\tt #1}}
\NewDocumentCommand\tool{m}{{\small\scshape #1}}

%% file: algos/finding-origin.tex
\begin{algorithmic}[1]
\footnotesize
\Function{findOrigins}{$CG$}
    \State $origins \gets emptyMapOfOrigins()$
    \ForAll {$edge_i \in CG.edges()$}
        \State{$target \gets edge_i.target()$}
        \State{$target_{type} \gets target.type()$}
        \ForAll{$parent_{type} \in target_{type}.parents()$}
            \If{$parent_{type}.isFirstDeclare(target.signature())$} 
                \State{$origins.put(target, parent_{type})$}
            \EndIf 
        \EndFor
    \EndFor
    
    \State \Return $origins$
\EndFunction
\end{algorithmic}

%% file: algos/localness.tex
\begin{algorithmic}[1]
\footnotesize
\Function{Categorize}{$method, CG$}
    \State $label \gets 0$
    \If{isDefinedInJava(method)}
        \State{\Return{$label$}}
    \EndIf
    \ForAll{$ edge_i \in CG.outgoingEdgesOf(method)$}
        \State{$ target_i \gets edge_i.target()$}
        \If{$!isDefinedInJava(target_i) $}
            \If{$ label<2 \And sameHierarchy(method, target_i)$}
                \State{$label \gets 1$}
            \Else
                \If{$areInSameProject(method, target_i)$}
                    \State{$label \gets 2$}
                \Else
                    \State{$label \gets 3$}
                    \State \textbf{break}
                \EndIf
            \EndIf
        \EndIf
    \EndFor
    \State{\Return{$label$}}
\EndFunction
\end{algorithmic}

%% file: algos/pruning.tex
\begin{algorithmic}[1]
\footnotesize
\Function{PruneCG}{$CG, exclusionList$}
    \State $pcg \gets new DirectedGraph()$ 
    \ForAll {$cs \in CG.getCallSites()$}
        \State $src \gets cs.getSource()$
        \State $tar \gets cs.getTarget()$
        \State $tarType \gets cs.getTargetType()$
        \State $tarSig \gets tarType.signature()$
        \State $excludedTypes \gets exclusionList.get(tarSig)$
        \If {$NotExcluded(ExcludedTypes, tarType)$}
            \State $pcg.addEdge(src, tar)$
        \EndIf
    \EndFor
    \State \Return{$pcg$}
\EndFunction

\Function{NotExcluded}{$exclusionList, targetType$}
    \ForAll {$type \in excludedTypes$}
        \If {$type.hasChild(targetType)$}
            \State \Return{False}
        \EndIf
    \EndFor
    \State \Return{True}
\EndFunction
\end{algorithmic}